\documentstyle[12pt,aasms4]{article}
\begin{document}

\newcommand{\lap}{$L_{38}^{-1/3}$}
\newcommand{\ergs}{\rm \su  erg \su s^{-1}}
\newcommand{\etal}{ {\it et al.}}
\newcommand{\porb}{ P_{orb} } 
\newcommand{\Po}{$ P_{orb} \su$}
\newcommand{\pdot}{$ \dot{P}_{orb} \,$}
\newcommand{\pot}{$ \dot{P}_{orb} / P_{orb} \su $}
\newcommand{\mm}{$ \dot{m}$ }
\newcommand{\mdot}{$ |\dot{m}|_{rad}$ }
\newcommand{\myr}{ \su M_{\odot} \su \rm yr^{-1}}
\newcommand{\msol}{\, M_{\odot}}
\newcommand{\ppp}{ \dot{P}_{-20} }
\newcommand{\cms}{ \rm \, cm^{-2} \, s^{-1} }
\newcommand{\pdott}{ \left( \frac{ \dot{P}/\dot{P}_o}{P_{1.6}^{3}} \right)}

\def\p{\phantom{1}}
\def\pmu{\mox{$^{-1}$}}
\def\ApJ{{\it Ap.\,J.\/}}
\def\ApJL{{\it Ap.\,J.\ (Letters)\/}}
\def\ApJS{{\it Astrophys.\,J.\ Supp.\/}}
\def\AJ{{\it Astron.\,J.\/}}
\def\AA{{\it Astr.\,Astrophys.\/}}
\def\AAL{{\it Astr.\,Astrophys.\ Letters\/}}
\def\AAS{{\it Astr.\,Astrophys.\ Suppl.\,Ser.\/}}
\def\MN{{\it Mon.\,Not.\,R.\,Astr.\,Soc.\/}}
\def\Na{{\it Nature \/}}
\def\SAIt{{\it Mem.\,Soc.\,Astron.\,It.\/}}
\def\kms{km^s$^{-1}$}
\def\sbu{mag^arcsec${{-2}$}}
\def\e{\mbox{e}}
\def\dex{\mbox{dex}}
\def\L{\mbox{${\cal L}$}}
\def\gte{\lower 0.5ex\hbox{${}\buildrel>\over\sim{}$}}
\def\lte{\lower 0.5ex\hbox{${}\buildrel<\over\sim{}$}}
\def\loe{\lower 0.6ex\hbox{${}\stackrel{<}{\sim}{}$}}
\def\goe{\lower 0.6ex\hbox{${}\stackrel{>}{\sim}{}$}}

\slugcomment{To appear in ApJ Letters}

\def\sgr{SGR~1900+14 }
\def\sgrp{SGR~1900+14}

\title{A  Giant Outburst from SGR 1900+14\\
Observed with the BeppoSAX
Gamma Ray Burst Monitor}

\author{M. Feroci\altaffilmark{1},
F. Frontera\altaffilmark{2,3},
E. Costa\altaffilmark{1},
L. Amati\altaffilmark{2},
M. Tavani\altaffilmark{4,5},
M. Rapisarda\altaffilmark{1,6}, 
M. Orlandini\altaffilmark{2}
}

\altaffiltext{1}{Istituto Astrofisica Spaziale, C.N.R., 
  Area di Ricerca Tor Vergata, Via Fosso del Cavaliere s.n.c.,
  00133 Roma, Italy}

\altaffiltext{2}{Istituto Tecnologie e Studio Radiazioni Extraterrestri,
CNR, Via Gobetti 101, 40129 Bologna, Italy}

\altaffiltext{3}{Dipartimento di Fisica, Universit\`a di Ferrara, Via Paradiso
 12, 44100 Ferrara, Italy}

\altaffiltext{4}{Istituto Fisica Cosmica,
CNR, Via Bassini 15, Milano, Italy}

\altaffiltext{5}{Columbia Astrophysics Laboratory,
Columbia University, New York, USA}

\altaffiltext{6}{ENEA - Sezione di Neutronica, Via Enrico Fermi 27,
 00044 Frascati, Italy}

\vskip 1in

\centerline{\it Astrophysical Journal Letters in press (1999)}

\begin{abstract}

We report the detection by the Gamma Ray Burst Monitor
onboard BeppoSAX of the strongest and longest ever 
detected outburst from \sgrp. 
%This event is characterized by a sudden rise
%($\sim 1$~s), a peak emission lasting $\sim 2$~s, and a
%fast quasi-exponential decay followed by a slower decay. 
Oscillations are detectable with a period of
$\sim 5.16$~s for the entire duration of the event ($\sim$300~s). 
The temporal analysis reveals also a remarkable
periodic substructure: after about 35~s from the event onset
each 5.16-s pulse shows a pattern of four subpulses and a dip,
each separated by $\sim1$~s.  
%No apparent relative phase variations or
%subpulse damping are observed in the oscillations for $\sim$100~s.
Significant spectral variation is detected during the
event and for each individual oscillation.
The first and most intense part of the outburst is quite 
hard, and similar to what previously detected
from the `March 5$^{th}$ event'. 
%(SGR~0525-66). 
A hard non-thermal spectral component persists for $\sim$200~s.
SGR~1900+14 was proposed to be a strongly  magnetized neutron star 
%with magnetic field 
($B\goe 10^{14}$~G) undergoing violent 
instabilities  by internal magnetic/crustal stresses. 
However, the onset of an apparent 1-s periodicity
within the 5.16-s pulsations and the observed spectral properties
show a complex behaviour that is not satisfactorily modelled yet.

\end{abstract}

Subject Headings: 
{gamma rays: bursts -- stars: individual (SGR~1900+14) -- 
stars: neutron -- X-rays: stars}

\section{Introduction}
Soft gamma-ray repeaters (SGRs) are compact objects undergoing spasmodic 
instabilities producing X-ray super-Eddington sub-second outbursts. 
%(sometimes in rapid succession, even 38 events in 350~s
%as in the case of \sgr on May 30th 1998, \cite{kouveliotou98a,hurley99b}).
%Recently, the accurate positioning of X-ray outbursts from
%SGR~1806-20 (\cite{murakami94}) confirmed that some SGRs
%are within Galactic supernova remnants (\cite{kulkarni93}).
%Also the remarkable source SGR~0525-66 that produced the
%`March 5th event' (showing the first evidence of a $\sim$8~s
%periodicity following a very intense initial pulse, \cite{mazets79a,cline80})
%is positionally coincident with the N49 supernova remnant
%in the Large Magellanic Cloud. 
The accurate positioning of SGR~0525-66 (the source of the
`March 5th event', showing the first evidence of a $\sim$8~s periodicity,
\cite{mazets79a,cline80}) and SGR~1806-20 (\cite{murakami94}) confirmed 
that some SGRs are within Galactic supernova remnants (\cite{kulkarni93}).
The combination of a relatively
long timescale oscillation (interpreted as rotation of a magnetized
compact object) and a lifetime comparable with that of the
associated remnant supported a model  based on a strongly
magnetized ($B \sim 10^{14}-10^{15}$~G) neutron star for
SGR~0525-66 (e.g., \cite{duncan92,paczynski92,thomson95}).
The recent detection of 7.47~s pulsations in the persistent flux from
SGR~1806-20 with period derivative
$8.3 \times 10^{-11} \rm s \, s^{-1}$  (\cite{kouveliotou98d})
%confirmed 
corroborated
this model for that repeater.

Among the four SGRs confirmed today, \sgr is peculiar for the
absence of a surrounding supernova remnant  
(however, it lies close to the less than $10^{4}$ years
old supernova remnant, G42.8+0.6, \cite{vasisht94,hurley99a,murakami99}).
Three outbursts of moderate intensity from
this source were first detected in 1979 (\cite{mazets79a}).
The source was detected  again in 1992 (\cite{kouveliotou93})
and in May 1998 (\cite{kouveliotou98a,hurley99a}).
ASCA and RXTE observations 
%after the May 1998 activity
revealed a 5.16~s periodicity in the quiescent 
X--ray (2--10 and 2--20 keV, respectively) emission  with period derivative 
$\sim 1 \times 10^{-10} \rm s \, s^{-1}$ (\cite{hurley99b,kouveliotou99}).
On 1998 August 27 a very strong outburst from
\sgr was detected by Konus-Wind, Ulysses and BeppoSAX
(\cite{cline98,hurley99a,feroci98}). The 5.16~s periodicity was 
unambiguously detected during the outburst. 
Transient radio emission was also detected (\cite{frail99})
one week after the event.
%: a previously unknown radio source
%appeared in the error box of \sgr on 1998, September 3 
%and faded away in a week.
These observational elements support an interpretation
of these phenomena in the framework of the {\it magnetar} 
model (\cite{thomson95}). 
However, in this paper we present data 
of the August 27 outburst from \sgr that are not yet 
explained by the current magnetar model.

\section{GRBM Observation}

The strong outburst from \sgr triggered the 
Gamma-Ray Burst Monitor (GRBM, \cite{frontera97,feroci97})
onboard the BeppoSAX satellite on 1998 August 27 10:22:15.7 UT.
The GRBM experiment consists of 4 CsI(Na) detection units, 10 mm thick,
with a geometric area of $\sim$1100 cm$^{2}$.
%The energy resolution of the each unit is 20\% at 280~keV. 
The data available from the GRBM include 1-s ratemeters 
from each of the 4 units in two partially overlapping
energy ranges (40--700~keV and $>$ 100~keV). 
Assuming no significant contribution 
from the spectral region above 700~keV 
(where the detector efficiency is $<$10\%)
we derive lightcurves in the two separate bands 40--100 and 
100--700~keV (\cite{amati97}). 
(The correctness of the peak fluxes given below is subject to 
this assumption.)
A dedicated memory allocated to the GRBM 
%for high time resolution data allows to 
records, for each trigger, the 40-700~keV 
lightcurve of the event from 8~s before to 98~s after the trigger, with a 
minimum time resolution of 7.8125~ms (9-bit counters). 
%for the 8~s pre-trigger (9-bit counters),
%0.489~ms for 10~s after the trigger (6-bit counters), 
%and 7.8125~ms in the following 88~s (10-bit counters). 
%Among the GRBM housekeeping data, 
240-channel energy spectra 
for each unit are continuously available, integrated over fixed
time intervals of 128~s, regardless of triggers. 

The event from \sgr was detected 
%with a high signal-to-noise ratio 
by the 4 GRBM detection units.
%, both in the main energy range (40--700~keV) and in the
%harder energy range (above 100 keV).   
The event occurred at an elevation angle of $48^{\circ}$ with respect 
to the GRBM equatorial plane, at an azimuthal angle of
$29^{\circ}$ with respect to the unit \#1 axis, and $61^{\circ}$ 
with respect to the unit \#4.  
The average effective area of unit \#1 at this direction varies from
$\sim$56~cm$^{2}$ at 60~keV to $\sim$365~cm$^{2}$ at 280~keV.
For the analysis presented here we
used only the signal from unit \#1 for which we have 
the highest signal-to-noise ratio. 
%Also, this unit is less affected by the energy-dependent 
%shadowing effect of satellite structures and 
%the its response matrix is more accurate than for unit \#4. 
%The count rates in units \#2 and \#3 are strongly affected by 
%scattering effects.
The count rates in the other units are strongly affected by
scattering effects and energy-dependent obscuration by
the satellite structures.

The estimated event duration is about 300~s. The total net counts 
were 1,007,699 in 40--700~keV and 301,235 above 100~keV. 
The net peak count rate in 40--700~keV was 147,812 counts s$^{-1}$ 
(average background $\sim$880 counts s$^{-1}$) 
and 67,025 counts s$^{-1}$
in the $>$100~keV band (average background $\sim$1,000 counts s$^{-1}$).
These values replace those reported in Feroci et~al. (1998) 
being corrected for dead-time and 
end-scale effects. In particular, we corrected our 1-s light curve
in the 40--700~keV range for one counter recycle (one recycle implies
an additive contribution of 65535 counts s$^{-1}$), 
by comparing the 1-s resolution data with our high time resolution
data (see below) and with the data from Ulysses (\cite{hurley99a}). 
This correction is what we consider the most likely to be applied, but
some uncertainty remains, and we present lower limits
to the peak flux and fluence. 
The 40--700~keV peak flux is detected 1~s after the pulse rise
and is $> 2.10 \times 10^{-4}$ erg~cm$^{-2}$~s$^{-1}$.
The measured fluence in the same band is $>1.5 \times 10^{-3}$ erg~cm$^{-2}$.
Assuming a source distance of 5 kpc (\cite{hurley99c}) and isotropic emission,
we obtain a luminosity
of $6 \times 10^{41}$ erg~s$^{-1}$ at the peak and a luminosity 
of $\sim3 \times 10^{40}$ erg~s$^{-1}$ in the subsequent $\sim$70~s. 
The total energy detected by the GRBM in the 40--700~keV band corresponds to
$\sim5 \times 10^{42}$ ergs. 
These flux values are obtained assuming a power law spectrum whose index 
is computed on the 1-s timescale from the two GRBM energy bands.

In Fig.~1 we show the 1-s background-subtracted lightcurve (top panel), 
together with the 7.8-ms lightcurve (rebinned at 31.25-ms, bottom panel, a) 
for the time interval when it is available and not affected by saturation.
%High time resolution data are only available for the first 98~s. 
%After that time, only a 1-s resolved lightcurve is available.         
In fact, given the intensity of the event, the counters of the high-resolution
data were saturated 
%(and therefore recycled several times) 
at the event peak. 
For this reason we cannot determine  the event risetime for timescales 
shorter than 1~s. 
The 5.16-s periodicity
%, first reported by Hurley et~al. (1999c), 
is clearly detectable in the 1-s resolved 
lightcurve overimposed to the general decay for the entire duration of 
the event. After a rapid decay during the first 2~s
(faster in the 100-700~keV range than in 40-100~keV), 
the decay can be approximated with two exponential laws, with
time constant $\tau\sim$5~s for the first $\sim15$~s and 
$\tau\sim$80~s for the subsequent decay.
The modulation of the oscillation with a $\sim$32~s period is
due to the sampling effect of a 5.16-s period at a 1~Hz frequency. 
The 31.25~ms lightcurve shows that, in addition to the 5.16-s periodicity, 
$\sim35$~s after the beginning of the event the 5.16-s pulse
is composed of 4 pulses and a dip 
equally separated in time by approximately 1~s (see next section).
In the high-resolution lightcurve we see no statistically significant
evidence for the precursor reported by Hurley et al. (1999a)
(in the 25--150~keV band, $\sim$0.4~s before the sharp rise).
Given the comparable statistical and temporal quality of our data
in the 40--700~keV, we are led to the conclusion that 
the spectrum of the precursor appears to be much softer than the rest 
of the emission.

\section{Temporal Analysis}

Fig.~2 shows the Power Spectral Density (PSD) for two portions 
of the 7.8-ms lightcurve. 
The PSD of the first part (37339.11--37371.11~s UT)
is shown in the top panel and clearly
demonstrates the presence of the 5.16-s pulsation since the beginning 
of the event. 
In the bottom panel of Fig.~2 we show the PSD for the time interval 
37369.13--37433.69~s UT, when the repetitive 1-s pattern is cleary set.
The latter PSD shows the presence
of the fundamental oscillation at 0.194~Hz, and of its harmonics at
0.389, 0.580, 0.775, 0.969 and 1.161~Hz. Higher orders
harmonics (up to n=19) are not visible in the plot due to scale compression.
%We computed the frequency of the main oscillation by using the 19$^{th}$
%order harmonic at 3.679044~Hz to increase the accuracy.
Our best estimate of the main pulsation 
(computed by using the 19$^{th}$ order harmonic at 3.679044~Hz 
to increase the accuracy) is $(0.1936\pm0.0008)$~Hz
(the baricenter correction is negligible), 
consistent with the more accurate ASCA measurement of 0.19384~Hz. 
(Hereafter the quoted errors correspond to a 90\% confidence level.) 
What is remarkable in the PSD shown in the bottom panel of Fig.~2 
is the strong enhancement of the power associated with the
{\it n}=5 harmonic at 0.969~Hz.
This is basically a measurement of the 1-s periodic structure 
within the 5.16-s pulse.
We have also searched for the period of the repetitive pattern
with no {\it a priori} assumptions: the maximum $\chi^2$ is obtained for
a period of $(1.03\pm0.03)$~s, that corresponds to 1/5 
of the fundamental period. 

Using the 1-s resolution data we have detected 
the fundamental oscillation at 0.194~Hz in the PSD 
obtained separately for the 40--100 and 100--700~keV energy
bands. A test for a phase-shift in these two energy bands using a 
cross-correlation analysis gave negative results.

\section{Spectral Analysis}

\subsection{Time-averaged Analysis}

The available 240-channel energy spectra for this event have 
initial and final times, with respect to the trigger time, as follows:
0--67~s (interval A), 68--195~s (interval B) and 
196--323~s (interval C).
We fitted to these spectra the phenomenologically
established\footnote{
We note that the OTTB model is purely phenomenological
and most likely without a physical basis for an
optically thick radiating medium
(see, for instance, \cite{fenimore94}). We use it here
for simplicity, and to relate our results to other
reports quoted in the SGR literature.}
spectral law for SGRs: an optically thin thermal bremsstrahlung 
%(OTTB, e.g. \cite{mazets79b,kouveliotou93,hurley95}).
(OTTB, e.g. \cite{mazets79b,kouveliotou93}).
A 10\% systematic error has been added to the statistical error 
to account for the calibration uncertainties at the
large off-axis direction.

The spectrum in interval A cannot be fitted 
with a simple OTTB law, showing 
an excess at high energies (similarly to what reported for the beginning of 
the March 5$^{th}$ event, e.g., \cite{mazets82,fenimore96}).
This is likely due to the strong spectral evolution as it appears 
from the 1-s ratemeters (see below), where we note a very hard
spectrum during the first seconds, becoming much softer soon after the peak.  
Assuming different spectral components for these two fractions
of interval A, we fitted the energy spectrum with the 
sum of an OTTB and of a single power law (PL, $I(E) \propto E^{-\alpha})$.
The best fit parameters are kT=$(31.2\pm2.5)$~keV and photon index  
$\alpha = 1.47\pm0.16$ (reduced $\chi^2$=0.705 with 186 d.o.f.).
However, the spectrum can be equally fitted with the sum
of a PL and a blackbody (BB), with $\alpha = 1.71\pm0.15$, and 
BB temperature kT=$(16.4\pm0.8)$~keV (reduced $\chi^2$=0.74).
%We have also tried to fit two OTTB laws,
%two power laws and a broken power law (that usually fits time-averaged 
%spectra of classical gamma-ray bursts) but the fits are unacceptable.
In interval B 
%from the ratemeter data we do not deduce strong spectral variation, therefore 
we first used a simple OTTB law. 
%However, 
The best fit to the spectrum has a kT=$(34.2\pm1.2)$~keV
and reduced $\chi^2$=1.996 (75 d.o.f.). We noted in the residuals the
existence of a hard excess that was likely biasing upward
the determination of kT. Thence, we  
added a PL component to the fit, and the best-fit parameters are
kT=$(27.6\pm1.9)$~keV and $\alpha = 4.5\pm0.2$ (reduced $\chi^2$=1.360,
73 d.o.f.).
Substituting the OTTB with a BB gives a reduced $\chi^2$=1.424 (73 d.o.f.) 
with kT=$(15.5\pm1.0)$~keV and $\alpha = 4.5\pm0.1$. 
For interval C we obtain a satisfactory OTTB fit
with kT=$(28.9\pm1.4)$~keV (reduced $\chi^2$=1.06, 71 d.o.f.). 
Fig.~3 shows the 3 energy spectra and best fit models.

\subsection{Time-resolved Analysis}

We used the 2-channel, 1-s ratemeters to study
short timescales spectral variation. We show our results
in terms of an hardness ratio (HR, counts in 100--700~keV
divided by those in 40--100~keV) and of the equivalent 
temperature kT of an OTTB law. 
The first 2~s can be described by an extraordinary high 
kT=3200~keV evolving to 60~keV, or with a value
of HR=4.06 decreasing to HR=0.75, respectively.
(We note that the peak values of kT and HR might be 
biased upward by the approximation mentioned in section 2).
% and the real value could be significantly smaller.) 
The subsequent evolution is shown in Fig.~1 (bottom panel) in the time 
interval for which the high time resolution data are available.
The HR and kT vary with the same $\sim$5-s periodicity as
in the lightcurve, with the hardness of the pulse
increasing almost linearly with phase, in a ramp-like 
fashion, with distinct maxima correlated with the dips in the lightcurve.
The global trend of the HR is a softening of the spectrum.
%Alternatively, we have also described the spectral evolution using a PL,
%deriving a variation of the photon index with time. We find that the
%first two seconds the 2-channel spectrum can be approximated by a
%photon index $\alpha$ = 0.7, evolving to about 3 in the next two seconds
%and then varying from 4 to 5 in the subsequent $\sim$100~s
%and between 5 and 6 in the rest of the event.  

\section{Discussion}

The huge event from SGR~1900+14 strikingly resembles 
the famous 1979 March 5$^{th}$ event. The main similar characteristics are:
(a) their long duration with respect to usual SGRs bursts;
(b) the impressive periodic pattern overimposed to a smooth exponential decay;
(c) the short initial hard peak.
% with the subsequent evolution to a softer spectrum.
%;(d) the spectral shape of the soft tail.
On the other hand, the August 27$^{th}$ event shows peculiarities not
reported for the March 5$^{th}$: 
(a) a much longer first peak ($\sim$1~s vs. $\sim$150~ms); 
(b) a hard spectral component persisting for $200\sim300$~s;  
(c) a strong evolution of the pulse shape, leading 
to a 1-s sub-pulse periodicity;
(d) a periodic spectral variation.
%(e) a much lower intrinsic luminosity both in the first pulse and
%in the soft tail.

Our results show a complex behavior of SGR~1900+14 
in outburst that can be resumed as follows.
(1) A clear determination of the 5.16~s periodicity 
{\it from the beginning of the outburst}. We note that this feature was not
detected by Ulysses (\cite{hurley99a}), implying that it
is related to the hard photons,
%the capability to detect the periodicity during this part of the outburst
%is related to the bandpass, 
similarly to the non-detection with the GRBM of the precursor 
detected by Ulysses. 
%It is also interesting that we detect
Our period determination is consistent (within an uncertainty of 0.02~s)
with the ASCA measurement, 
implying that such a large outburst caused a glitch with 
$\Delta P/P < 3.1 \times 10^{-3}$.
(2) We see a clear transition after $\sim$35~s in the shape of the 5.16-s pulse
profile. At the beginning the pulse is composed of two broad 
pulses, with relative intensity gradually changing with time.
At the 6$^{th}$ pulse from the onset of the outburst the two peaks become
sharper and gradually two additional narrow peaks appear. A striking
feature of these 4 pulses is their equal mutual distance in time and
their phase stability within the 5.16-s period. This feature 
appears in the PSD as a net enhancement of the n=5 harmonic with respect
to the others, corresponding to a periodicity of  $(1.03\pm0.03)$~s.
It looks like the 5.16-s periodicity in this part of the outburst
is basically due to the lack/occultation of a 5$^{th}$ peak that one
could have expected in correspondence of the 5.16-s dip. 
%The occurrence of such a 5-th peak would have resulted in the 
%disappearance of the 5.16-s periodicity.
(3) The determination of a complex spectral evolution showing a very
hard initial outburst pulse followed by a softer emission
%(parametrized with an OTTB with kT$\simeq30$~keV) 
that is
strongly modulated in agreement with the periodic pulse structure.
We notice a `see-saw' behaviour of the HR, with the highest hardness
in correspondence of the dips of the lightcurve. If the dip
would be due to an occultation this would imply that a 
persistent hard component is present in the emitted spectrum.
A hard component is observed in the time-averaged spectra  
as a PL component additional to the single OTTB
(see Fig.~3, top and middle panel).
(4) Finally, we note that the decay of the event lightcurve as detected
by the GRBM is well approximated by a double exponential law and
cannot be described by a power law 
%$I(t) \propto t^{-\alpha}$
as derived for the Ulysses lightcurve (\cite{hurley99a}). Also this 
difference is likely related to the different Ulysses-BeppoSAX/GRBM 
bandpasses. 

The emerging picture is that of a complex phenomenon, involving
a non-trivial response of a compact object, most likely
a highly magnetized neutron star with a structured emitting
region, to a major explosive event.
Our point no. 2 is worth of special attention. The emitting
region surrounding the compact object is first subject to a
violent readjustment following the initial strongly
super-Eddington outburst (possibly related to a relativistic
ejection producing the transient radio emission, \cite{frail99}). 
Subsequent evolution
and settling of the 1-s periodic feature reveals a highly
structured emitting region with the excitation of higher order,
possibly resonant oscillating modes. 
%In particular, the strong enhancement of the fifth harmonic that we
%observed 
This feature is unprecedented among SGRs strong outburst detections.
Trapping in a magnetosphere of emitting `blobs' (\cite{thomson95}) 
co-rotating with the surface of a neutron star, or having their
proper oscillation mode, might explain
the harmonic content of the time variable power spectrum of
SGR~1900+14.

\section{Acknowledgements}
%We thank Enrico Massaro for useful discussions 
%and Kevin Hurley for discussing Ulysses data with us, for making 
%his manuscripts available to us prior to publication and for 
%careful reading of the manuscript.
We thank E. Massaro for useful discussions 
and K. Hurley for discussing Ulysses data with us
and for careful reading of the manuscript.

%\newpage
%
%\centerline{Figure Captions}
%
%\vskip .3in
%FIGURE 1:   
\begin{figure}
\figurenum{1}
\plotone{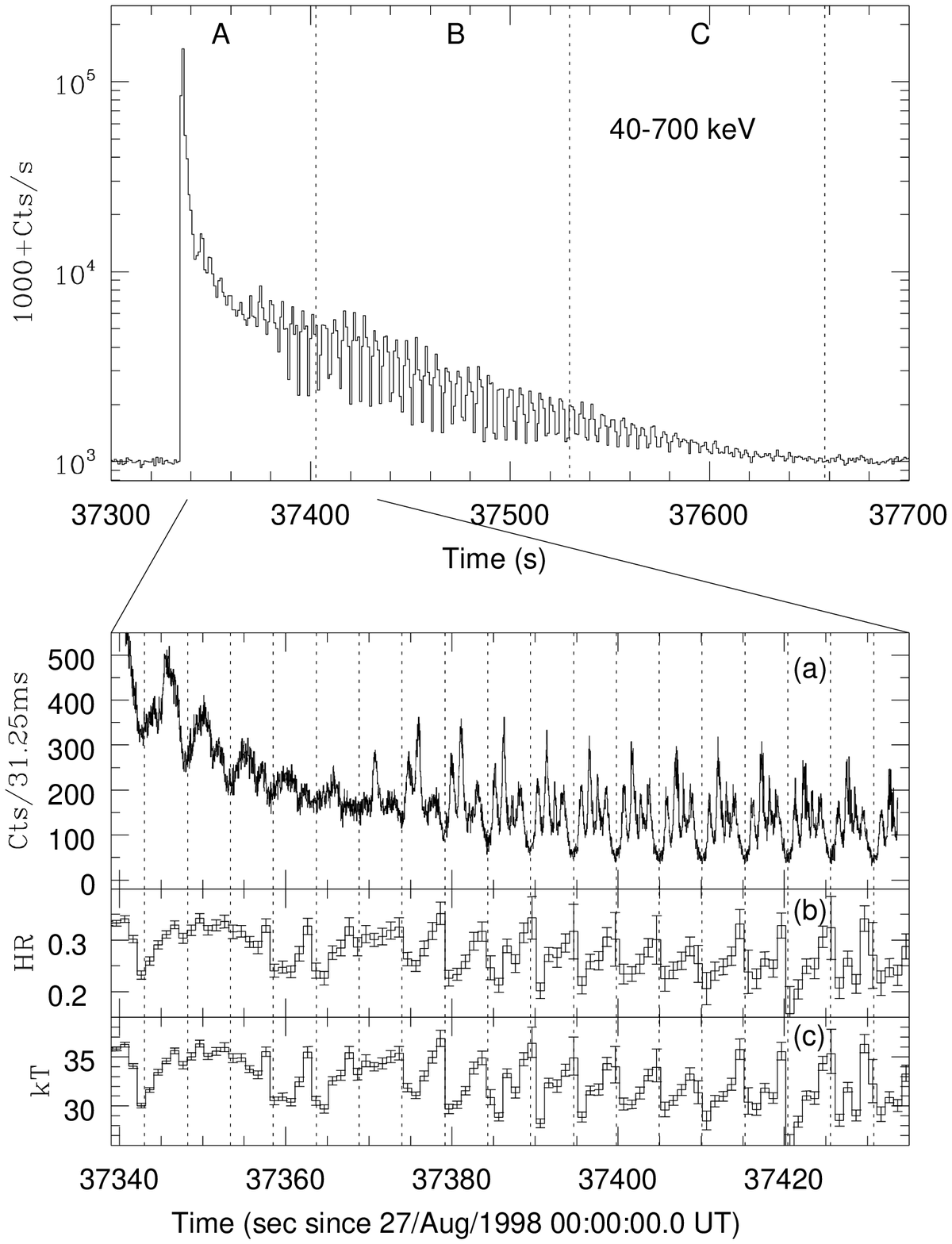}
\caption{
{\it Top}: 40--700~keV 1-s background subtracted light curve of the 
event (a value of 1000 counts/s has been added for display purposes).
The dashed vertical lines define the intervals A, B and C for which
we have the time averaged energy spectra.
{\it Bottom}: (a) high resolution lightcurve (rebinned at 31.25~ms) 
of the available portion of the event ($\sim$100~s);
(b) 1-s resolved spectral evolution described in terms of the simple
hardness ratio HR=(100--700~keV)/(40--100~keV) and
(c) in terms of an equivalent kT of an optically thin thermal 
bremsstrahlung. 1-$\sigma$ errors are shown. 
The vertical dotted lines are spaced by one period  
according to Hurley et al. (1999a). 
}
\end{figure}

%\vskip .3in
%FIGURE 2:  
\begin{figure}
\figurenum{2}
\plotone{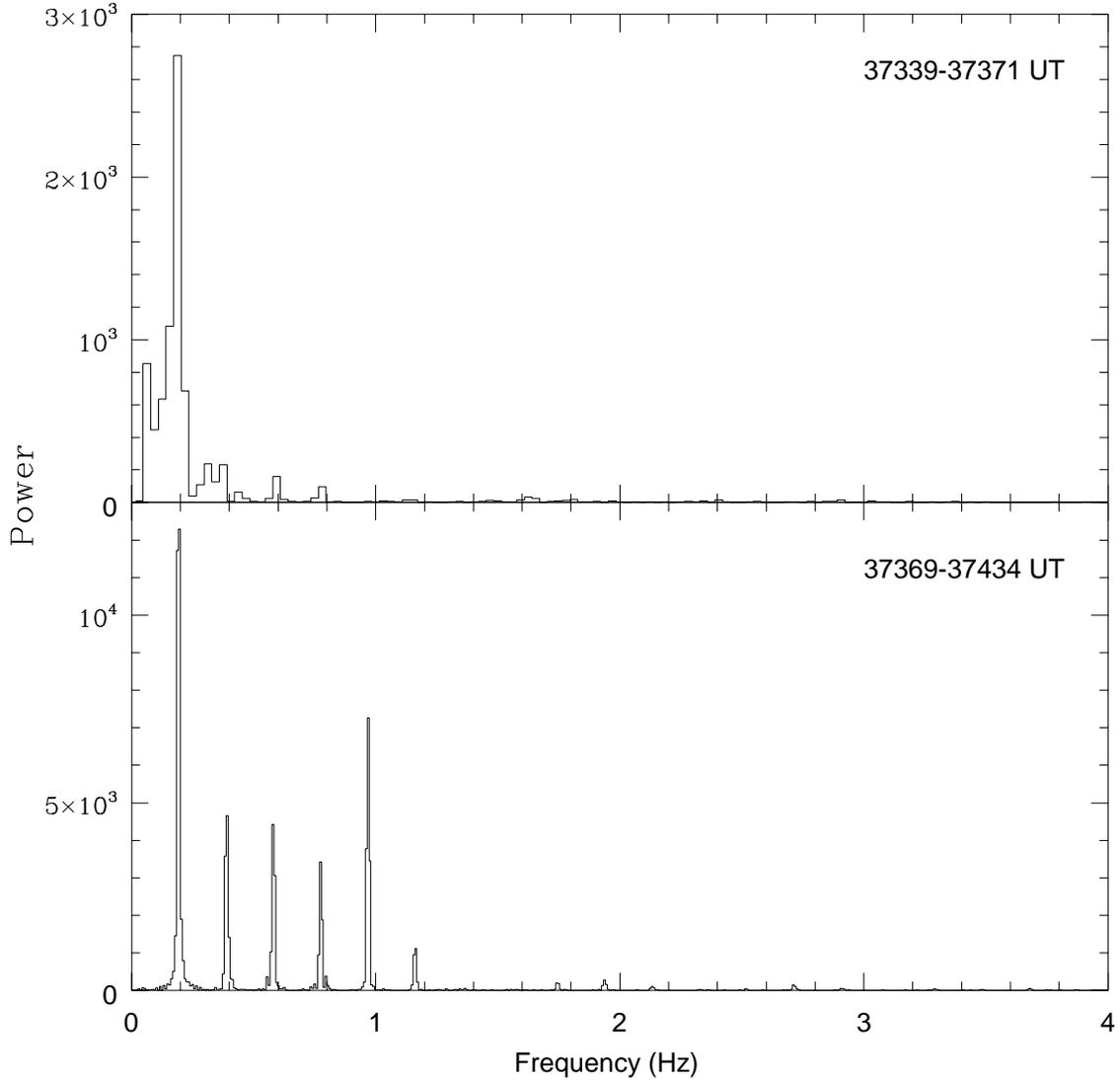}
\caption{
Power spectral density of the 7.8~ms lightcurve in the 40--700~keV range. 
{\it Top}: the time interval from 37339.11--37371.11 UT, demonstrating
the presence of the 5.16-s pulsation from the beginning of the event.
{\it Bottom}: the last 64~s of the high resolution light curve 
(from 37369.13 to 37433.68 UT), when the 1-s subpulse structure is clearly set.
(A four-degrees polynomial detrending procedure has been applied.)
%in order to subtract the global decay trend of the lightcurve.
}
\end{figure}

%\vskip .3in
%FIGURE 3: 
\begin{figure}
\figurenum{3}
\plotone{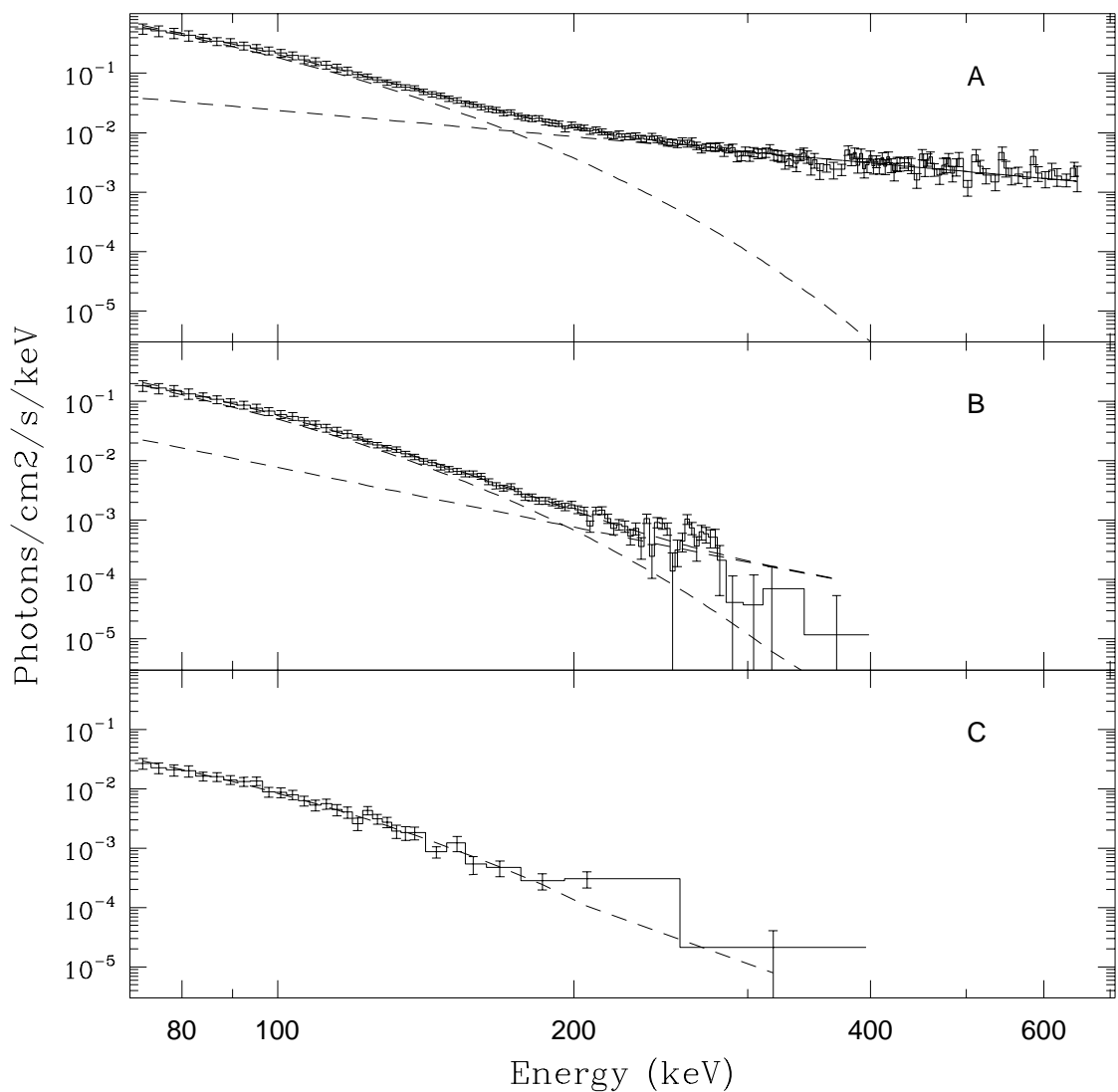}
\caption{
 Fit to the GRBM spectrum (the individual spectral components are
 shown as dashed lines) in the energy range 70--650~keV
 (A) and 70--400~keV (B and C) with a model including:
 (A) OTTB (kT=31~keV) plus PL ($\alpha$=1.47)
% (A) An Optically Thin Thermal Bremsstrahlung (OTTB, kT=31~keV) 
% plus a Power Law (PL, $\alpha$=1.47) 
 for the time interval from 0 to 67~s
 with respect to the trigger time; (B) OTTB plus PL with 
 kT=28~keV and $\alpha$=4.5 for the time interval from 68 to 195~s;
 (C) OTTB with kT=29~keV for the time interval from 196 to 323~s.
}
\end{figure}


\begin{thebibliography}{aa}


\bibitem[Amati et~al.\ 1997]{amati97}
Amati, L. et~al. 1997, SPIE Proceedings, 3114, 176

\bibitem[Cline et~al. 1980]{cline80}
Cline, T.L., et al., 1980, ApJ, 237, L1

\bibitem[Cline et~al.\ 1998]{cline98}
Cline, T.L., Mazets, E.P., and Golenetskii, S.V. 1998,
IAU Circ. 7002

\bibitem[Duncan \& Thompson 1992]{duncan92}
Duncan, R.C. \& Thompson, C., 1992, ApJ, 392, L9

\bibitem[Fenimore et~al.\ 1994]{fenimore94}
Fenimore, E.E., Laros, J.G., and Ulmer, A., 1994, ApJ, 432, 742

\bibitem[Fenimore et~al.\ 1996]{fenimore96}
Fenimore, E.E., et al., 1996, ApJ, 460, 964

\bibitem[Feroci et~al.\ 1997]{feroci97}
Feroci, M. et~al. 1997, SPIE Proceedings, 3114, 186

\bibitem[Feroci et~al.\ 1998]{feroci98}
Feroci, M. et~al. 1998, IAU Circ. 7005 

\bibitem[Frail et~al.\ 1999]{frail99}
Frail, D., Kulkarni, S., and Bloom, J., 1999, Nature {\it in press}
 
\bibitem[Frontera et~al.\ 1997]{frontera97}
Frontera, F., 
et al., 
%Costa, E., Dal~Fiume, D., Feroci, M., Nicastro, L., Orlandini,
%M., Palazzi, E., and Zavattini, G. 
1997, A\&AS, 122, 357

%\bibitem[Hurley 1995]{hurley95}
%Hurley, K., 1995b, in {\it X-Ray Binaries},
%eds. W. Lewin, J. van Paradijs, \& E. van den Heuvel,
%(Cambridge Astrophysics Series 26), p.532

%\bibitem[Hurley et~al.\ 1998a]{hurley98a}
%Hurley, K. et~al. 1998a, IAU Circ. 6929 
%
%\bibitem[Hurley et~al.\ 1998b]{hurley98c}
%Hurley, K. et~al. 1998b, IAU Circ. 7001

\bibitem[Hurley et~al.\ 1999a]{hurley99a}
Hurley, K. et~al. 1999a, Nature, 397, 41

\bibitem[Hurley et~al.\ 1999b]{hurley99b}
Hurley, K. et~al. 1999b, ApJ, 510, L107

\bibitem[Hurley et~al.\ 1999c]{hurley99c}
Hurley, K. et~al. 1999c, ApJ, 510, L111
 
\bibitem[Kouveliotou et~al.\ 1993]{kouveliotou93}
Kouveliotou, C., et~al. 1993, Nature, 362, 728
 
%\bibitem[Kouveliotou et~al.\ 1998a]{kouveliotou98c}
%Kouveliotou, C., et al., 1998a, GCN Circ. 94

\bibitem[Kouveliotou et~al.\ 1998a]{kouveliotou98d}
Kouveliotou, C., et~al. 1998a, Nature, 393, 235

\bibitem[Kouveliotou et~al.\ 1998b]{kouveliotou98a}
Kouveliotou, C. et~al. 1998b, IAU Circ. 6929 

\bibitem[Kouveliotou et~al.\ 1999]{kouveliotou99}
Kouveliotou et~al. 1999, ApJ, 510, L115

\bibitem[Kulkarni \& Frail 1993]{kulkarni93}
Kulkarni, S. \& Frail, D., 1993, Nature, 365, 33

\bibitem[Mazets et~al.\ 1979a]{mazets79a}
Mazets, E.P., Golenetskii, S.V., and Gur'yan, Yu.A. 1979a,
Soviet Astron. Lett., 5(No.6),343

\bibitem[Mazets et~al.\ 1979b]{mazets79b}
Mazets, E.P., et al., 1979b, Nature, 282, 587

\bibitem[Mazets et~al.\ 1982]{mazets82}
Mazets, E.P., Golenetskii, S.V., Gur'yan, Yu.A., Ilyinskii, V., 1982,
Astrophys. Space Sci. 84, 173

\bibitem[Murakami et~al.\ 1994]{murakami94}
Murakami, T.. et al., 1994, Nature, 36, 129

\bibitem[Murakami et~al.\ 1999]{murakami99}
Murakami, T. et~al. 1999, ApJ, 510, L122

\bibitem[Paczy\`nski 1992]{paczynski92}
Paczy\`nski, B., 1992, Acta Astron., 42, 145

\bibitem[Thompson \& Duncan 1995]{thomson95}
Thompson, C., and Duncan, R.C., 1995, MNRAS, 275, 255

\bibitem[Vasisht et~al.\ 1994]{vasisht94}
Vasisht, G., et al., 1994, ApJ, 431, L35

\end{thebibliography}
\end{document}